\documentstyle[emulateapj,onecolfloat,psfig]{article}

\newcommand{\ApJ}{Astrophys. J.}

\newcommand{\lin}{{\rm lin}}

\def\sun{\hbox{$\odot$}}

\newcommand{\vir}{{\rm vir}}

\newlength{\tskip}
\newlength{\colwidth}

\newcommand{\beq}{\begin{equation}}
\newcommand{\eeq}{\end{equation}}
\newcommand{\beqa}{\begin{eqnarray}}
\newcommand{\eeqa}{\end{eqnarray}}
\newcommand{\bn}{\hat{\bf n}}

\newcommand{\Ylmn}{Y_{l}^{m}}

\begin{document}
\twocolumn[
\title{Kinetic Sunyaev-Zel'dovich effect from halo rotation}
\author{Asantha Cooray$^1,$\altaffilmark{2}, Xuelei Chen$^3$}
\affil{$^1$Theoretical Astrophysics,
California Institute of Technology, Pasadena,  CA 91125.\\
$^3$Institute for  Theoretical Physics, Kohn Hall, University of California, Santa Barbara, CA 93106.\\
E-mail: asante@hyde.uchicago.edu, xuelei@itp.ucsb.edu}

%\preprint{NSF-ITP-02-09}

\begin{abstract}
We discuss the kinetic Sunyaev-Zel'dovich (SZ) contribution to
cosmic microwave background (CMB) temperature fluctuations
due to coherent rotational velocity component of electrons within
halos. This effect produces a distinct dipole-like temperature 
distribution across the the cluster, and provides a 
promising way to measure the 
angular momentum distribution of gas inside clusters. Information
obtained from such a measurement may provide new
insights to the origin and evolution of angular momentum in
hierarchical structure formation theory.
For typical, well relaxed, clusters of mass a few times 10$^{14}$ M$_{\sun}$,
the peak fluctuation is of the order a few $\mu$K, 
depending on the rotational
velocity and the inclination angle of the rotational axis. For clusters
which had underwent a recent merger, the contribution to temperature fluctuations could be even larger.
This dipole signature is similar to the one produced by lensed CMB
towards galaxy clusters, though the
lensing contribution spans a larger angular extent than the one due
to rotational scattering as the former depends on the gradient of the
cluster potential. Since the lensing contribution towards
clusters are aligned with the large scale CMB gradient, when higher 
resolution observations towards clusters are combined with a wide 
field CMB map, these two effects can be separated. An additional, but
less important, source of confusion is the dipolar pattern produced by the moving-lens
effect involving, again, the gradient of the cluster potential and the
transverse velocity. The angular power spectrum of temperature  anisotropies produced by the halo rotation
is expected to be smaller than  those due to the thermal SZ and peculiar velocity kinetic SZ effects. 
\end{abstract}

]

\altaffiltext{2}{Sherman Fairchild Senior Research Fellow}

\section{Introduction}
In recent years, stimulated by the great advances in the experimental front,
theoretical study of secondary anisotropies in the cosmic microwave
background (CMB) temperature has been carried out to much
lengthy detail. At large angular scales, the acoustic peak structure
of CMB  provides detailed information related to cosmological
parameters (\cite{Eisetal99} 1999), while at smaller angular scales, the CMB
anisotropies contain important information related to local large scale
structure (e.g., \cite{Coo01} 2001). 

Among the secondary contributions important at small angular scales, 
the thermal Sunyaev-Zel'dovich (SZ; \cite{SunZel80} 1980) effect
due to the inverse-Compton scattering of CMB
photons by hot electrons in galaxy clusters is now well studied
both experimentally (\cite{Caretal96} 1996; \cite{Jonetal93} 1993) 
and theoretically  (\cite{Spretal01} 2001; \cite{Coo00} 2000;
\cite{KomKit99} 1999; \cite{MolBir00} 2000; \cite{Coo01} 2001).
Since the thermal SZ effect bears a distinct
spectral signature, with multi-frequency data, it is possible to
separate its contribution from others, allowing a detailed study of its
properties (\cite{Cooetal00} 2000). Another contribution,
the kinetic SZ effect, comes from the {\it peculiar} motion of
electrons in the rest frame of CMB photons (\cite{SunZel80} 1980).
This effect is also known as the Ostriker-Vishniac
effect in the linear regime of density fluctuations (OV;
\cite{OstVis86} 1986; \cite{Vis87} 1987) and has been studied in
detail analytically (e.g., \cite{Hu00} 2000; \cite{Coo01} 2001) and numerically
(\cite{Spretal01} 2001; \cite{MaFry01} 2001) during recent years given the possible observational
measurement. At these small scales, other important contributions
include weak gravitational lensing of CMB photons (e.g.,
\cite{SelZal00} 2000) and contributions via scattering through inhomogeneous reionization
scenarios (e.g., \cite{Aghetal96} 1996).

In this paper, we study a possible contribution to the CMB anisotropy
that results from the rotation of clusters. The effect is similar to the
kinetic SZ effect, however, the electron motion, relative to the rest
frame of CMB, is now due to the rotational velocity of the 
associated halo and not the usual line of
sight peculiar motion. With the rotational axis aligned across the
line of sight, the effect has a useful observational signature 
involving a distinct negative and positive temperature 
decrement towards a given halo. For spherically symmetric halos, 
this dipolar pattern is symmetric and its peak magnitude and 
the extent allows a measurement of the rotational
velocity when combined with other observations towards galaxy clusters that
probe the baryon distribution.
  
As far as we know, in previous analytical calculations of kinetic SZ effect,
this contribution has not been included.
However, we note that this rotational contribution
should be present in the total kinetic SZ contribution measured in
numerical simulations, even though in previous studies no
attempt has been made to separate this rotational velocity of gas 
from the peculiar motions. Though this effect
may not be well represented  in current simulations due to issues
associated with resolution, the general agreement between current analytical 
theory and numerical models (\cite{MaFry01} 2001) of the kinetic SZ effect then
suggests that this effect may be statistically small. Nevertheless, we
find that the effect is important for certain well aligned individual
clusters. The rotational contribution to the kinetic SZ effect also
has several interesting applications as we detail here.

In general, the collisionless components of galaxy clusters 
are supported by velocity dispersion, and the gas component by pressure.
The cluster temperature is related to these through virial
equilibrium. Still, some rotation in the cluster is generically
expected, with a magnitude of at least a few percent of the circular
velocity. Such a rotation produce a dipole pattern in a kinetic SZ map of
the cluster, which could then be used to measure 
the angular momentum of the cluster gas. As we shall discuss in \S~2, 
such measurements could shed light on the origin and evolution of
angular momentum in the hierarchical structure formation scenario,
which is at present one of the outstanding problems in models of 
galaxy formation.

The galaxy cluster rotations may also have an important role in massive 
cooling flows:  the associated energy transport of baryons within 
the halo may provide a solution to one of the well known 
problems in astrophysics today involving a lack of low 
temperature gas in cooling flows as observed by Chandra 
and XMM recently (\cite{Bohetal01} 2001). 

Extending our investigation to a distribution of galaxy clusters, we also
calculate the angular power spectrum of anisotropies produced by
randomly oriented rotating halos. We find that the halo rotation make
less contribution to the angular power spectrum than the thermal SZ effect
and the kinetic effect due to halo peculiar velocities. Nevertheless, 
this rotation-generated dipole could also complicate efforts to
detect a similar dipole-like pattern in temperature
produced by gravitational lensing of CMB photons by 
galaxy clusters (\cite{SelZal00}  2000).
Here, it was suggested that the dipole signature related
to lensing can be used to extract the lensed CMB contribution, and
thus, some aspect of the lensing potential. 
We suggest that such a study may be contaminated 
by the rotational kinetic SZ signal. However,
when high resolution CMB observations towards clusters are combined with a
wide-field CMB map, the dipolar patterns produced by the rotational
kinetic SZ effect and the gravitational lensing effect 
can be separated out based
on the fact that the dipole due to lensing is aligned with 
the large scale CMB gradient. The large scale CMB gradient 
typically spans several tens of arcminutes, while a typical 
cluster spans at most few arcminutes on the sky.
In some cases, the large scale CMB gradient is negligible, 
for example if the cluster happens to be 
on top of a CMB hot or cold spot, instead where the background
temperature changes. In such a favorable scenario,
the rotational contribution may provide the only dominant dipolar pattern.

This paper is organized as follows: 
in \S~2, we briefly review the current understanding and modeling
of the cluster rotation. In \S~3, we calculate the rotational halo
contribution to CMB anisotropies due to an individual cluster. 
In \S~4, we discuss angular power spectrum of temperature fluctuations and
conclude with a discussion in \S~5.  Though we present our derivations
for arbitrary cosmology, for the illustration of
our results, we take a $\Lambda$CDM cosmology with parameters
$\Omega_m=0.35$ for the matter density, $\Omega_b=0.05$ for the baryon
density, $\Omega_\Lambda=0.65$ for the cosmological constant, $h=0.65$
for the Hubble constant, and a scale-invariant spectrum
of primordial fluctuations, normalized to galaxy cluster abundances
($\sigma_8=0.9$; see, \cite{ViaLid99} 1999).
and consistent with COBE
normalization of \cite{BunWhi97} (1997).
For the linear power
spectrum, we use the fitting function for the transfer function given by
\cite{EisHu99} (1999).

\begin{figure*}[tb]
\centerline{\psfig{file=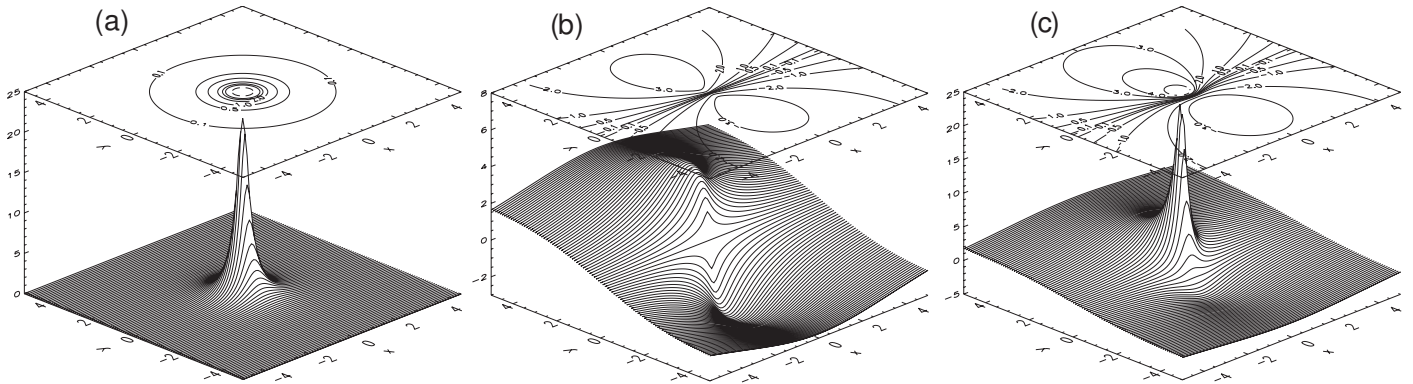,width=\textwidth,angle=0}}
\caption{Temperature fluctuations due to galaxy clusters: (a) kinetic SZ
effect involving peculiar motion,
(b) lensing of CMB primary temperature fluctuations, and (c)
the total contribution from kinetic SZ, lensing and rotational
velocity. The total contribution shows an asymmetric dipolar pattern with
a sharp rise towards the center. 
We use the same cluster as shown in figure~\ref{fig:rotational}.}
\label{fig:secondary}
\end{figure*}

\section{The Rotation of dark halo and cluster gas}

The angular momentum plays an important role in the formation of 
structures, yet at present, our understanding of its origin and 
evolution is very incomplete. A widely accepted model for the 
origin is that the angular momentum is induced by 
tidal torques from the surrounding matter (\cite{Hol53} 1953;
\cite{Pee69} 1969; \cite{Dor70} 1970; \cite{Whi84} 1984).
During the linear growth stage, the angular momentum of matter increases
until over dense regions reach their maximum size and then turn
around, collapsing in to halos (\cite{CatThe96} 1996; \cite{Sugetal00}
2000). This predicts  a log normal distribution for the spin parameter, 
as seen in N-body simulations (e.g., \cite{BarEfs87} 1987; \cite{HeaPea88}
1988; \cite{Zuretal88} 1988) with the right order of
magnitude. In galaxy size halos, the
baryons cool and contract while conserving
angular momentum and,  eventually, form disks
(e.g.,\cite{Mes63} 1963; \cite{WhiRee78} 1978; \cite{FalEfs80} 1980).
 
There are a number of problems, however, confronting the 
tidal torque theory. The predicted rotation is about three times higher than simulations
indicate (e.g., \cite{BarEfs87} 1987), and the prediction on the direction of spin 
has large errors (e.g., \cite{LeePan00} 2000). Recently, an alternative suggestion
was made in which the angular momentum
were acquired from the orbital angular momentum of merging
satellites (\cite{Maletal02} 2002; \cite{Vitetal01} 2001). 
While a complete understanding is still lacking, the distribution of 
angular momentum could be ``measured'' in 
high resolution numerical simulations. One study found that the 
distribution of angular momentum among halos could be 
characterized as log normal, with typical rotation velocity about 5\% of circular velocity.
For most halos, there is a ``universal'' spin profile 
consistent with solid body rotation. The solid-body rotation, however,
saturates at large values for the angular momentum.
The spatial distribution of angular momentum in 
most halos, 80\% of the sample, tend to be cylindrical and
well-aligned while  the halo spin is almost independent of its
mass  and does not evolve with redshift except after major
mergers (\cite{Buletal01} 2001). The existence of this  
``universal spin profile'' is still being debated; while there is a
fraction of halos whose rotation is not well aligned, a greater 
percentage of such mis-aligned halos were found in a different study
(\cite{vdBetal02} 2002).  Still, we note that these two studies do agree
on the total amount of the angular momentum. Since in observed galaxies
such misaligned halos are not very common, in \cite{vdBetal02} (2002) it
is speculated that the negative angular momentum material combine with
positive ones to form bulges. In the context of galaxy clusters, following similar 
arguments, one should expect to find these material form cD galaxies in the
center.

The problem become much more complicated, however, when baryons are taken into account.
A well known problem in galaxy formation theory is the ``angular
momentum catastrophe'': overcooling gas looses significant amount
of angular momentum and contracts to a disk which is an order of 
magnitude smaller than observed (\cite{NavBen91} 1991; \cite{NavWhi94}
1994; \cite{NavSte00} 2000). The overcooling of 
gas could be prevented by supernova feedback (e.g., \cite{Weietal98}
1998) which is some what uncertain. 

The observational study of cluster rotations may provide interesting clues for solving
some of these problems.  If angular momentum of cluster gas can be
measured, it may provide a potentially ``cleaner'' test of our
understanding on the origin and evolution of angular momentum. 
Since CMB observations can probe, potentially, to a higher redshift without
the $(1+z)^4$ brightness dimming encountered in classical X-ray studies,
one can expect to study not only nearby clusters but also a population
of high redshift clusters. Such an approach allows a survey of angular
momenta for the understanding of its evolution.

At present, there is limited information on gas rotational velocities
in galaxy clusters since high resolution simulations of clusters are
limited to dark matter and not gas. Inspired by the study of halo 
angular momentum distribution in \cite{Buletal01} (2001) for dark
matter,  to illustrate our calculations, we shall assume  that 
(i) the dark matter have a universal angular momentum profile in each halo, 
which is very close to solid body rotation;
(ii) the distribution of the spin parameter satisfies a log normal
distribution; and (iii) electrons have the same
angular momentum distribution as the dark matter.

As discussed above, the hypothesis (i) involving the existence of a 
dark halo universal spin profile is still debated. 
We expect this issue to be resolved, eventually, with the 
improvements in numerical simulations.
In this {\it analytical} study, we shall tentatively adopt it
as a working hypothesis. This  allows us to make an estimate 
of the magnitude of the  signal in CMB temperature fluctuations. 
We suggest that once high resolution
maps of the kinetic SZ effect become available from next
generation CMB experiments, this hypothesis will
be tested observationally. The simulations 
of \cite{vdBetal02} (2002) and \cite{Maletal02} (2002) indicate that
the size of the spin parameter of dark matter and gas are similar,  
though they do not necessarily align in the same direction. Thus,
our third assumption should be right, even if the gas does not
corotate with dark matter.

\begin{figure}[t]
\centerline{\psfig{file=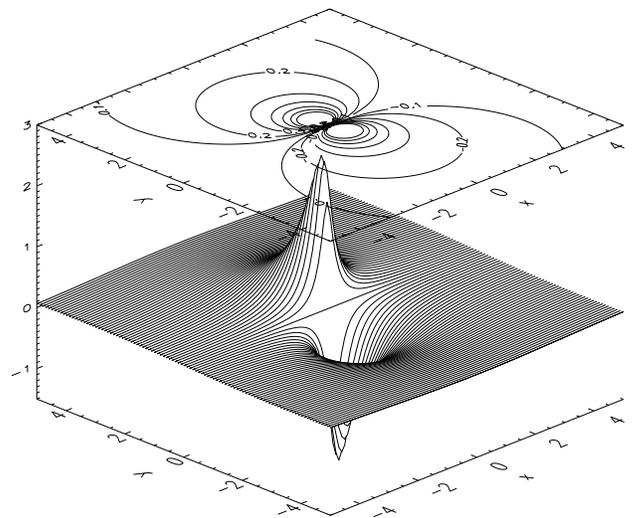,width=4.3in,angle=90}}
\caption{
Contribution to temperature fluctuations through halo
rotation for a cluster of mass $5 \times 10^{14}$ M$_{\sun}$ at a
redshift of 0.5. The temperature fluctuations produce a distinct
dipolar-like pattern on the  sky with a maximum of $\sim$ 2.5 $\mu$K.
Here, rotational axis is perpendicular to the line of sight and
$x$ and $y$ coordinates are in terms of the
scale radius of the cluster, based on the NFW profile.}
\label{fig:rotational} 
\end{figure}

\section{Calculation}

We can write the temperature fluctuations due to the kinetic
SZ effect as
\begin{equation}
\frac{\Delta T}{T}(\bn) = \int d\chi \sigma_T n_e e^{-\tau} \bn \cdot
{\bf v} \, ,
\end{equation}
where ${\bf v}$ is the velocity field of electrons with number density
$n_e$, $\tau$ is the optical depth to scattering and $\sigma_T$ is the
Thomson cross-section.

The peculiar velocity kinetic SZ effect resulting from the large scale structure has
now been well  studied (e.g., \cite{Hu00} 2000; \cite{MaFry01} 2001).
In analytical calculations, one usually describes the velocity field of
electrons through the peculiar motions associated with large scale
structure bulk flows. In this respect, calculations usually involve the linear
theory description of the velocity field, which is
applicable at large scales. At
non linear scales, in addition to peculiar motion, the velocity field
of electrons in galaxy clusters is likely to include
an additional component due to rotation.

Assuming that the dark matter motion could also be described by the 
universal spin profile (\cite{Buletal01} 2001), we adopt the simplifying
assumption of {\it solid-body} rotation for each cluster as discussed
in \S~2. This allows us to
prescribe halo rotations through a constant angular velocity,
$\omega$, for each cluster. One can easily modify our calculation for other
rotational profiles for cluster gas, such as a  rotational 
component that is a function of the radius, $\omega(r)$.
We take an angular momentum profile for baryons  that follow
dark matter; this is consistent with simulations even if the gas
motion does not follow that of the dark matter exactly \cite{vdBetal02} (2002). 
This assumption should be valid at large radii and beyond a typical cluster
core radius. In the inner-most region, rotational velocities of electrons may differ from
that of the dark matter due to effects related to pressure cut-off, 
additional effects resulting from the presence of significant 
magnetic fields, and massive cooling flows. 
Note that, currently, we do not have
detailed information related to cluster gas rotation, 
both observationally and numerically.
The limited information we have is consistent 
with our assumptions: for example,
\cite{Efs88} (1988) found that the ratio of rotational velocity 
to velocity dispersion in galaxy clusters at the radii 
where overdensity is about 500 times the background is 
about 0.15, which, as we find later, is somewhat higher
than estimates based on the assumption that gas rotate with dark
matter, $\sim 0.12$, though consistent with uncertainties in
observational data and numerical studies. 

For an individual cluster at a redshift $z$ with an angular
diameter distance $d_c$, we can write the temperature fluctuation
as an integral of the electron density, $n_e(r)$, weighted by
the rotational velocity component, $\omega r \cos(\alpha)$,
along the line of sight. Introducing the fact that the line of sight
velocity due to rotation is proportional to sine of the inclination
angle of the rotational axis with respect to the observer, $i$, we write
\begin{equation}
\frac{\Delta T}{T}(\theta,\phi) = \sigma_T e^{-\tau} \eta(\theta)
\cos \phi \sin i \,
\label{eqn:rotation}
\end{equation}
where
\begin{equation}
\eta(\theta) =  \int_{d_c\theta}^{R_{\vir}} \frac{2 r
dr}{\sqrt{r^2 - d_c^2\theta^2}}
n_e(r) \omega d_c \theta .
\label{eqn:cluster}
\end{equation}
Here, $\theta$ is the line of sight angle relative to the cluster
center and $\phi$ is an azimuthal angle measured relative to an axis
perpendicular to the spin axis in the plane of the sky.
In simplifying, we have introduced the fact that 
the angle between the rotational velocity and line of sight, $\alpha$, is such
that $\cos \alpha = d_c \theta/r$.
In Eq.~\ref{eqn:cluster}, $R_\vir$ is the cluster virial radius and we
model the galaxy cluster dark matter distribution as prescribed
by \cite{Navetal96} (1996; NFW) with a scale radius
$r_s$,
\begin{equation}
\rho_\delta(r) = \frac{\rho_s}{(r/r_s)(1+r/r_s)^{2}} \, .
\end{equation}
Note that the  density profile can be integrated and related to total dark
matter mass of the halo within $R_\vir$
\begin{equation}
M_\delta =  4 \pi \rho_s r_s^3 \left[ \log(1+c) - \frac{c}{1+c}\right]
\label{eqn:deltamass}
\end{equation}
where the concentration, $c$, is $R_\vir/r_s$. Alternatively, spherical
collapse tells us $M = 4 \pi r_v^3 \Delta(z) \rho_b/3$, where $\Delta(z)$ is
the overdensity of collapse and $\rho_b$ is the background matter density
today. By equating these two expressions, one can
eliminate $\rho_s$ and describe the halo by its mass $M$ and
concentration $c$.

To describe the baryon distribution  in clusters, we make use of the hydrostatic
equilibrium  to calculate a  profile for the gas distribution
following \cite{Maketal98} (1998). The hydrostatic equilibrium is
a valid assumption given that current observations of halos, mainly galaxy clusters, suggest
the existence of regularity relations, such as size-temperature (e.g.,
\cite{MohEvr97} 1997), between physical properties of dark matter and
baryon distributions. We refer the reader to \cite{Coo00} (2000) for full
details. Using the hydrostatic equilibrium allows us to calculate the
baryon density profile, $\rho_g(r)$, within halos
\begin{equation}
\rho_g(r) = \rho_{g0} e^{-b} \left(1+\frac{r}{r_s}\right)^{br_s/r} \, ,
\label{eqn:gasprofile}
\end{equation}
where $b$ is a constant, for a given mass,
\begin{equation}
b = \frac{4 \pi G \mu m_p \rho_s r_s^3}{k_B T_e} \, ,
\label{eqn:b}
\end{equation}
with the Boltzmann constant, $k_B$ \cite{Maketal98} (1998). 
In general, the halos are described with virial temperatures
\begin{equation}
k_B T_e =  \frac{\gamma G \mu m_p M_\delta(r_v)}{3 r_v} \, ,
\label{eqn:virial}
\end{equation}
with $\gamma=3/2$ and $\mu=0.59$, corresponding to a hydrogen mass
fraction of 76\%. Since $r_v \propto M_\delta^{1/3}(1+z)^{-1}$ in
physical coordinates, $T_e \propto M^{2/3}(1+z)$.

To describe the halo rotations, we write the dimensionless spin
parameter $\lambda (= J \sqrt{E}/GM^{5/2})$ following \cite{Buletal01}  (2001)
as
\begin{equation}
\lambda = \frac{J}{2 V_c M_\vir R_\vir} \frac{\sqrt{cg(c)} }{f(c)} \,
\label{eqn:lambda}
\end{equation}
where the virial concentration for the NFW profile is
$c=R_\vir/r_s$, $J$ is the total angular momentum, and $V_c^2= G M_\vir/R_\vir$.
To relate angular velocity, $\omega$, to spin, we
integrate the NFW profile over a cluster to calculate $J$ and write
\begin{equation}
\omega = \frac{3 \lambda V_c c^2 f^2(c)}{R_\vir h(c) \sqrt{c g(c)}} \, .
\label{eqn:omega}
\end{equation}
The functions $f(c)$, $g(c)$ and
$h(c)$, in terms of the concentration, follows as
\begin{eqnarray}
f(c) &=& \ln(1+c) -\frac{c}{1+c}\nonumber \\
g(c) &=& 1-\frac{2\ln(1+c)}{1+c} -\frac{1}{(1+c)^2} \nonumber \\
h(c) &=& 3\ln(1+c) +\frac{c(c^2-3c-6)}{2(1+c)} \, .
\end{eqnarray}
To set the normalization for the baryon profile, 
we take the baryon fraction in galaxy clusters to be the
universal baryon fraction such that $M_{\rm gas}/M_\vir =
\Omega_b/\Omega_m$ (see, \cite{Coo00} 2000 for details).
One can modify the fraction of baryons, as a function of halo mass, from
the universal assumption used here. We note that, though the kinetic
SZ effect is sensitive to the galaxy groups (\cite{Coo01} 2001) whose gas distribution
may be uncertain, the rotational contribution is skewed towards
massive clusters: at the virial radius, the velocity due to rotation
scales as $M_{\rm vir}^{1/3}$, for a sample of halos with the same concentration.
Thus, the calculations presented here are not fully dependent on the
highly uncertain gas distribution in smaller halos such as groups.

In \cite{Buletal01} (2001), the
probability distribution function for $\lambda$ was measured through
numerical simulations and was found to be well described by a log
normal distribution with a mean, $\bar{\lambda}$, of $0.042 \pm 0.006$
and a width, $\sigma_\lambda$ of $0.50 \pm 0.04$. We use the mean
value of $\lambda$ to calculate $\omega$ for individual clusters;
we also found results based on integrating the exact probability
distribution function to be consistent with results based on the
mean. Note that in simulations of \cite{Efs88} (1988), 
the spin parameter was found to be in the range 
of 0.04 to 0.06 with a mean of $\sim 0.05$, slightly above the estimate
by \cite{Buletal01} (2001).

\section{Individual clusters}
\label{sec:fisher}
In Fig.~\ref{fig:rotational}, we show the temperature fluctuation
produced by the rotational component for a typical cluster with mass
$5 \times 10^{14}$ M$_{\sun}$ at a redshift of 0.5. The maximal effect
is on the order of $\sim$ 2.5 $\mu$K with a sharp drop towards the center
of the cluster due to the decrease in the rotation velocity. As
shown, the effect leads to a distinct temperature distribution with a
dipole like pattern across clusters. Here, we have taken the cluster
rotational axis to be aligned perpendicular to the line of sight; as
it is clear, when the axis is aligned along the line of sight, there
is no resulting contribution to the SZ kinetic effect through
scattering. Additionally, we note that the halo rotations can only contribute to CMB
temperature fluctuations through scattering processes. In terms of gravitational
redshift related contributions, the momentum associated with the
rotational velocity does not contribute to the non-linear
integrated Sachs-Wolfe effect as the resulting time-derivatives of the
potential fluctuations are zero (\cite{Coo02} 2002).

The order of magnitude  of this rotational contribution can be
understood by estimating the rotational velocity where the effect peaks.
In Eq.~\ref{eqn:omega},
rotational velocity is $\omega \sim 3 \lambda V_c/R_\vir$ with functions
depending on the concentration in the order of a few ($\approx 2.4$
when $c=5$). Since the circular velocity for typical cluster is of
order $\sim$ 1500 km s$^{-1}$, with $R_\vir \sim$  Mpc and
$\bar{\lambda} \sim 0.04$, at typical inner radii of order $\sim 1/5
R_\vir$, we find velocities of order $\sim$ 36 km s$^{-1}$ and increases to about
$\sim 180$ km s$^{-1}$ at the outer radii of the cluster. Since, on average,
peculiar velocities for massive clusters are of order $\sim$ 300 km s$^{-1}$
in our fiducial $\Lambda$CDM cosmology (see, \cite{SheDia01}), the
rotational velocity, at typical core radii of clusters, is lower by a factor of $\sim$ 10,  
when compared with the peculiar velocity of the typical
cluster. Similarly,  at the outer radii, the ratio of rotational velocity to  velocity 
dispersion of the cluster is of order 0.12, 
consistent with the expectation of 0.15 by \cite{Efs88} (1988).

Furthermore, since the
kinetic SZ due to peculiar motion peaks in the center of the halo where
the density is highest, while the
rotational effect peaks away from the center, the 
difference between maximal peculiar
kinetic SZ and rotational kinetic SZ temperature fluctuations is
even greater. Note, however, each individual cluster has a 
different orientation
and magnitude of peculiar velocity and rotation, thus the velocity-to-rotation
ratio could vary a lot. In favorable cases where the peculiar velocity is
aligned mostly across the line of sight, the rotational contribution
is important.

It should be noted that the above discussion applies to typical
clusters found in blind surveys. 
For initial attempt of detecting this effect, one may
observe more favorable cases where there are indications of significant
rotational velocity from other observations. For example, hints for
significant rotation may come from evidence for a recent merger, 
large ellipticity and bimodal distribution of cluster
galaxies. Indeed, there is some observational 
evidence (\cite{DupBre01} 2001) that in two nearby clusters, the Perseus cluster (Abell 426) 
and Centaurus cluster (Abell 3526), the gas is rotating with velocity
of a few 1000 km s$^{-1}$; This rotational velocity is much greater than the typical
case we have assumed and produces few tens of $\mu$K signal in the CMB
data instead of the few $\mu$K signal. In such nearby merging
clusters, the rotational contribution may be
detectable with instruments  that are just becoming available, such as
the BOLOCAM (Lange, private communication).

In \cite{DupBre01} (2001), the evidence for gas rotation comes from the
redshift gradient of X-ray emission lines across the cluster surface.
The use of X-ray lines allow another method to probe gas motions in clusters.
Unfortunately, the spectral resolution of the X-ray telescopes
are limited and makes it difficult to search for smaller, and more 
typical, rotational velocities in relaxed
clusters.   Furthermore, for more distant clusters, the photon flux is lower and the detection is almost
impossible. This is not a problem for observations associated with
fluctuations imprinted on CMB as the sensitivity does not decrease
significantly with redshift\footnote{We note that, however,
one need improving angular resolution to probe more distant clusters.}.
Also, we note that the sensitivity of X-ray emission is proportional to $n_e^2$, while
the contribution to CMB only depends on the electron density, $n_e$.
Thus, we expect the SZ kinetic signal due to rotation to fall off less
rapidly than the X-ray emission when one looks at larger radii from
the cluster center. 

In Fig.~\ref{fig:secondary}, we show the
kinetic SZ effect towards the same cluster due to the peculiar motion
and the contribution resulting from the lensed CMB towards the same
cluster. The latter contribution is sensitive to the gradient of the
dark matter potential of the cluster along the large scale CMB
gradient. In this illustration, we  haven taken the CMB gradient
to be the rms value with 13 $\mu$K arcmin$^{-1}$ following
\cite{SelZal00} (2000). Previously, it was suggested that the lensed
CMB contribution can be extracted based on its dipole like
signature. Given the fact that the rotational contribution also leads
to a similar pattern, any temperature distribution with a dipole
pattern across a cluster cannot easily be prescribed to the lensing
effect. 

There is also another source of confusion resulting from the dipolar
pattern towards clusters produced by the moving-lens effect
(\cite{MolBir00} 2000) resulting from the non-linear Rees-Sciama
effect (\cite{ReeSci68} 1968; see Cooray 2002 for a recent
calculation). The signal here is again depends on the gradient of the
cluster potential, but, weighted by the transverse velocity across the
line of sight. The resulting temperature fluctuations are again, at
most, a few $\mu$K.  The dipolar pattern produced by this effect is
aligned with the direction of the motion across the sky, which need
not be same as the direction of the large scale CMB gradient which is
lensed. Therefore, in principle, it is possible to separate these two
confusing signals from each other, as well as, for the rotational contribution.

The possibility for separation could be understood through Figs.~\ref{fig:rotational} and
\ref{fig:secondary}; the dipole signature associated with the
rotational scattering is limited to the inner region of the cluster
while the lensing effect, as well as the moving-lens effect, due to their dependence on the gradient of the
dark halo potential, covers a much larger extent. Also, the 
dipole due to rotation need not lie in the same direction as the background gradient
of the primary CMB fluctuations as the rotational axis of halos may
be aligned differently. In Fig.~\ref{fig:secondary}, we have not included the
dominant thermal SZ contribution since it can be separated  from
other contributions reliably if multi-frequency data are available.

Note that the lensing effect is aligned with the large scale 
gradient of CMB; thus, when higher resolution CMB observations 
of clusters are combined with a wide-field CMB map, one 
can extract the lensing contribution by noting the direction 
of the large scale gradient of the CMB across the cluster. 
By using the shape of the gradient, one can the  extract the dipolar
pattern due to the moving-lens effect.
The separation of lensing effects and the rotational contribution from each other and from dominant 
kinetic SZ effect can be effectively carried out in Fourier  space through various filtering schemes (see,
discussion in \cite{SelZal00} 2000) with certain filters effectively aligned based on
a priori knowledge. In certain favorable cases, the only dipolar 
contribution may be due to the rotational component as the cluster 
may sit on top  of a hot or a cold spot of the CMB temperature 
fluctuations instead of a regime where temperature is varying and may
not have an appreciable transverse component of velocity. 
The former scenario is more likely to be  possible as the 
temperature fluctuations in CMB have characteristic scales 
of $\sim$ square degree while the extent of galaxy clusters 
at high redshifts are of the order few arcminutes. In such cases, the
two dipolar patterns when separated will give the transverse and
rotational components of the velocity, while the dominant kinetic
effect will give the peculiar velocity; using higher resolution CMB
data alone, it is, thus, possible to map the true three-dimensional
velocity distribution of galaxy clusters.

\begin{figure}
\centerline{\psfig{file=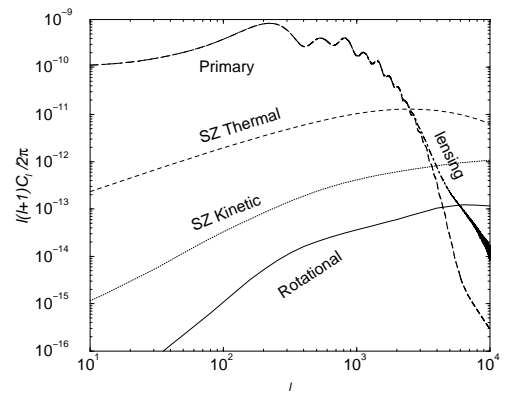,width=0.35\textwidth,angle=-90}}
\caption{
Angular power spectra of temperature anisotropies due to
rotational scattering (solid line) compared to contributions from
other effects towards clusters: SZ thermal (dashed line) and SZ kinetic
(dotted line) primary (long dashed line) and lensed (dot-dashed line)
CMB anisotropies.}
\label{fig:power}
\end{figure}

\section{Angular Power Spectrum}

In order to calculate the angular power spectrum of  temperature
fluctuations resulting from the rotational velocities, we make use of
a cluster population with a mass function, $dn(M,z)/dM$,
given by the Press-Schechter (PS; \cite{PreSch74} 1974) theory.
Following previous approaches to
calculate the angular power spectra of the thermal and kinetic SZ
effects (see \cite{Coo00} 2000; \cite{KomKit99} 1999; \cite{MolBir00}
2001; \cite{Coo01} 2001  and references therein),
we write the angular power spectrum
of temperature fluctuations as
\begin{equation}
C_l^{1h} = \int dz \frac{dV}{dz} \int dM \frac{dn(M,z)}{dM}
\frac{2}{3} G_l^2 \, .
\end{equation}
Here, the factor of $2/3$ accounts for the random distribution of
rotational inclination angles.
The function $G_l$ accounts for the spherical harmonic transformation of a
cluster temperature profile, with a mass $M$ and at a redshift $z$,
produced by the rotational effect. We can write $G_l$ as
\begin{equation}
G_l^2 = \frac{1}{2l+1} \sum_{m=-l}^{l} |a_{lm}|^2 \, ,
\end{equation}
where $a_{lm}$ are multipole moments following $\int d\bn T(\bn)\Ylmn {}^*(\bn)$.
Due to the dependence of azimuthal angle, $\cos \phi$, in Eq.~\ref{eqn:rotation},
only $m=-1$ and +1 contributes to the power spectrum.
To calculate multipole moments, we follow
\cite{MolBir00} (2000) and perform a two-dimensional integration of the cluster profile
over associated Legendre polynomials, $P_l^m(\cos \theta)$
\begin{eqnarray}
a_{l1} &=& - \sqrt{\frac{2l+1}{4\pi} \frac{(l-1)!}{(l+1)!}} \int dA_c
\eta(\theta) P_l^1(\cos \theta) \cos^2 \phi \, , 
\end{eqnarray}
Since $a_{l1}=-a_{l-1}$, due to the reality of the temperature
fluctuation field,   and using $P_l^1(\cos \theta) =
(l+1/2)J_1[(l+1/2)\theta]$ with first order Bessel function of the
first kind given by $J_1$, we can write, in the large $l$ limit,
\begin{equation}
G_l^2 = \frac{1}{2} \frac{F_l^2}{4\pi}
\end{equation}
with the Legendre coefficients given by an
equivalent 2-dimensional Fourier transform  for a
temperature distribution with an azimuthal dependence
\begin{equation}
F_l = 2\pi \int_0^{\theta_\vir} d\theta \eta(\theta) J_1(l\theta) \, .
\end{equation}

Since halos themselves are clustered in their spatial distribution,
with a 3-dimensional linear power spectrum of $P(k)$,
there is an additional contribution at large angular scales following
\begin{eqnarray}
C_l^{2h} &=& \int dz \frac{dV}{dz} P^\lin\left[k=\frac{l}{d_A};z\right]
\nonumber \\
&\times& \frac{2}{3} \left[ \int dM \frac{dn(M,z)}{dM} b(M,z)
G_l\right]^2 \, ,
\end{eqnarray}
where halo biasing, $b(M,z)$, relative to linear theory can be obtained
according to the
prescription of \cite{Moetal97} (1997). This contribution is the so-called
{\it two-halo} term under the context of halo approach to non-linear
clustering, while
the former $C_l^{1h}$ is the {\it single-halo} contribution sometimes called the Poisson
term (see, e.g., \cite{Sel00} 2000; and for a recent review,
\cite{CooShe02} 2002).

Considering the statistical detection of the rotational contribution,
we computed the angular power spectrum of temperature
anisotropies.
Our results are summarized in Fig.~\ref{fig:power}. The solid line
shows the angular power spectrum due to the rotational component while
the curved labeled ``kinetic SZ'' is the usual contribution associated
with bulk motion of halos (see \cite{Coo01}  2001 for calculational
details). For comparison, we also show the thermal SZ contribution and
the lensed CMB contribution. In general, anisotropy power spectrum
resulting from rotational contribution is smaller than current
estimates of SZ thermal and SZ kinetic power. This conclusion,
however, is subjected to the rotational velocity description. Moreover,
our spherical halo model may also underestimate this effect because
it is quite plausible that fast rotating halos are non-spherical and thus
have higher density away from the center. We
encourage further numerical work, especially in hydrodynamic
simulations, to establish the full extent to which electron rotational
velocities within halos may be important as a source of secondary
fluctuations in the CMB temperature. Understanding the rotational
component is necessary to
establish the fully non-linear kinetic SZ effect contributions to CMB
temperature fluctuations.

\section{Summary}
\label{sec:discussion}

We presented an aspect of  the kinetic Sunyaev-Zel'dovich contribution to
cosmic microwave background temperature fluctuations
due to the coherent rotational velocity component of 
electrons within halos, instead of the usual line of 
sight peculiar velocity. This contribution has not 
been explicitly considered in prior works and we believe 
the observational signature due to rotation may provide a useful probe of the
physics of gas rotations within halos. 

For typical clusters of mass a few times 10$^{14}$ M$_{\sun}$,
this rotational scattering effect produces
a distinct dipole-like temperature distribution with a peak
fluctuation of the order a few $\mu$K, depending on the rotational
velocity and the inclination angle of the rotation axis. For an initial 
detection, one may also look for clusters with indications of such
significant rotational velocity through evidences such as a recent
merger, a bimodal distribution of  galaxies, a large ellipticity or X-ray emission line gradients.
In such case of recent mergers, the rotation velocity could be few tens of
times greater than the typical cases we have considered allowing
detection with instruments such as BOLOCAM that are just becoming available.
For more typical cases involving relaxed clusters, the contribution
due to rotational velocities can eventually be
probed with future higher resolution CMB observations 
such as those planned with the CARMA array and the South Pole 
Telescope (J. Carlstrom, private communication).

This dipole signature is similar to
the one produced by lensed CMB towards galaxy clusters, though the
lensing contribution spans a larger angular extent than the one due
to rotational scattering. Since the lensing contribution towards 
clusters are aligned with the large scale CMB gradient, when higher 
resolution observations towards clusters are combined with a 
wide field CMB map, the gravitational lensing dipole can be
separated out from the dipole 
pattern towards the cluster due to the rotational contribution.
The angular power spectrum of temperature
anisotropies produced by the halo rotation
is below the fluctuations power associated with the thermal SZ and
peculiar velocity kinetic SZ effects. 

The rotational kinetic SZ effect allows a useful probe of 
the angular momentum of gas in massive galaxy clusters, 
which is not easily measurable from other observational techniques.
This could provide us useful information on the evolution of clusters
and galaxies. Understanding the rotation of cluster baryons may also be
be helpful in the context of 
massive cooling flows where there is now a well known 
problem in astrophysics with the lack of low temperature gas.

\acknowledgments    
We thank the organizers of the 2001 Santa Fe
Cosmology and Large Scale Structure Workshop where this work was initiated.
AC is supported at Chicago by NASA grant NAG5-10840 and at Caltech by
the Sherman Fairchild foundation and by the DOE under DE-FG03-92ER40701. 
XC is supported at OSU by the DOE under
grant DE-FG02-91ER40690, and at ITP/UCSB by the NSF under grants
PHY99-07949.


\begin{thebibliography}{99}

\bibitem[Aghanim et al]{Aghetal96}
        Aghanim, N., Desert, F. X., Puget, J. L., \& Gispert, R. 1996,
A\&A, 311, 1

\bibitem[Barnes \& Efstathiou]{BarEfs87}
	Barnes, J. \& Efstathiou, G. 1987,  ApJ, 319, 575 

\bibitem[B\"ohringer et al]{Bohetal01}
B\"ohringer,  H., Matsushita, K.,  Churazov, E.,
Ikebe, Y., and Chen, Y. 2001, A\&A submitted (astro-ph/0111112).


\bibitem[Bullock et al.]{Buletal01}
        Bullock, J. S., Dekel, A., Kolatt, T. S., et al. 2001, ApJ,
555, 240

\bibitem[Bunn \& White]{BunWhi97}
       Bunn E. F., \&  White, M. 1997, ApJ,480, 6 


\bibitem[Carlstrom et al.]{Caretal96}
        Carlstrom, J. E., Joy, M., Grego, L. 1996, ApJ, 456, L75

\bibitem[Catelan \& Theuns]{CatThe96}
	Catelan, P., \& Theuns, T. 1996, MNRAS, 282, 436

\bibitem[Cooray]{Coo01}
        Cooray, A. 2001, Phys. Rev. D.,  {\bf 64}, 063514  (2001).

%\bibitem[Cooray \& Hu]{CooHu01a}
%        Cooray, A., Hu, W. 2001a, ApJ in press, astro-ph/0004151
 
\bibitem[Cooray et al.]{Cooetal00}
        Cooray, A., Hu, W., Tegmark, M. 2000, ApJ, 540, 1

\bibitem[Cooray]{Coo00}
      Cooray, A., 2000, Pys. Rev. D., 62, 103506

\bibitem[Cooray]{Coo02}
Cooray, A. 2002, Phys. Rev. D. in press, astro-ph/0109162

\bibitem[Cooray \& Sheth]{CooShe02}
Cooray, A. \& Sheth, R. K. 2002, Phys. Rep. in press

 
%\bibitem[Cooray et al]{Cooetal00b}
%        Cooray, A., Hu, W., Miralda-Escud\'e, J. 2000b, ApJ 536, L9
 
%\bibitem[Dawson et al.]{Dawetal00}
%Dawson, K. S., Holzapfel, W. L., Carlstrom, J. E., Joy, M., LaRoque,
%S. J.,
%\& Reese, E. D., 2000, ApJ, submitted, astro-ph/0012151

\bibitem[Doroshkevich]{Dor70}
	Doroshkevich, A. G. 1970, Astrofizika, 6, 581 
 
\bibitem[Dupke \& Bregman]{DupBre01} 
Dupke, R. A. \& Bregman, J. N. 2001, ApJ, 547, 705

\bibitem[Efstathiou et al]{Efs88}
Efstathiou, G., Frenk,  C. S., White, S. D. M. \& Davis, M. 1988,  MNRAS, 235, 715

\bibitem[Eisenstein \& Hu]{EisHu99}
        Eisenstein, D.J. \& Hu, W. 1999, ApJ, 511, 5

\bibitem[Eisenstein et al]{Eisetal99}
Eisenstein, D. J., Hu, W., and Tegmark, M. 1999, ApJ, 518, 2.


\bibitem[Fall \& Efstathio]{FalEfs80}
		Fall, S. M., \& Efstathiou, G. 1980, MNRAS, 193, 189 



 
%\bibitem[Haiman et al.]{Haietal00}
%       Haiman, Z., Mohr, J. J., Holder, G. P. 2000, astro-ph/0002336

%\bibitem[Henry]{Hen00}
%        Henry, J. P. 2000, ApJ, 534, 565


\bibitem[Heavens \& Peacock]{HeaPea88}
 Heavens,  A. \& Peacock, J. A. 1988, MNRAS, 232, 339

\bibitem[Hoyle]{Hol53} 
	Hoyle, F. 1953, ApJ, 118, 513 


 
\bibitem[Hu]{Hu00}
        Hu, W. 2000, ApJ, 529, 12
 
\bibitem[Jones et al]{Jonetal93}
Jones, M. Saunders, R., Alexander, P., et al. 1993, Nature, 365, 320
 
\bibitem[Komatsu \& Kitayama]{KomKit99}
Komatsu, E. \& Kitayama, T., 1999, ApJ, 526, L1

\bibitem[Lee \& Pen]{LeePan00}
	 Lee, J.,  Pan, U.-L. 2000, ApJ, 532, L5 
 
\bibitem[Limber]{Lim54}
        Limber, D. 1954, ApJ, 119, 655
 
\bibitem[Ma \& Fry]{MaFry00}
        Ma, C.-P., Fry, J. N. 2000, ApJ, 538, L107

\bibitem[Ma \& Fry]{MaFry01}
        Ma, C.-P., Fry, J. N. 2001, preprint (astro-ph/0106342)
 
\bibitem[Makino et al.]{Maketal98}
        Makino, N., Sasaki, S., Suto, Y. 1998, ApJ, 497, 555

\bibitem[Maller et al]{Maletal02}
	Maller, A. H., Dekel, A.,  \& Somerville, R. S. 2002, MNRAS, 329, 423

\bibitem[Mestel]{Mes63} 
	Mestel, L. 1963, MNRAS, 126, 553 
 
\bibitem[Mo \& White]{MoWhi96}
        Mo, H. J., White, S. D. M. 1996, MNRAS, 282, 347
 
\bibitem[Mo et al.]{Moetal97}
        Mo, H. J., Jing, Y. P., White, S. D. M. 1997, MNRAS, 284, 189

\bibitem[Mohr \& Evrard]{MohEvr97}
         Mohr, J. J. \& Evrard, A. E. 1997,  ApJ, 491, 38
 
\bibitem[Molnar \& Birkinshaw]{MolBir00}
Molnar, S. M., Birkinshaw, M. 2000, ApJ,  537, 542
 
\bibitem[Navarro \& Benz]{NavBen91}
	Navarro, J. F. \& Benz, W. 1991, ApJ, 380, 320

\bibitem[Navarro \& White]{NavWhi94}
	Navarro, J. F. \& White, S. D. M. 1994, MNRAS, 267, 401

\bibitem[Navarro \& Steinmetz]{NavSte00}
	Navarro, J. F. \& Steinmetz, M. 2000, ApJ, 538, 477 

\bibitem[Navarro et al]{Navetal96}
        Navarro, J., Frenk, C., White, S. D. M., 1996, ApJ, 462, 563
[NFW]
 
\bibitem[Ostriker \& Vishniac]{OstVis86}
        Ostriker, J.P., \& Vishniac, E.T. 1986, Nature, 322, 804

 
 
%\bibitem[Peacock \& Dodds]{PeaDod96}
%        Peacock, J.A., Dodds, S.J. 1996, MNRAS, 280, L19
 
\bibitem[Peebles]{Pee69}
	Peebles, P. J. E. 1969, ApJ, 155, 393

        Peebles, P.J.E. 1980, The Large-Scale Structure of the
Universe,        (Princeton: Princeton Univ. Press)
 
%\bibitem[Peebles]{Pee93}
%        Peebles, P.J.E. 1993, Principles of Physical Cosmology,
%        (Princeton: Princeton Univ. Press)
 
\bibitem[Press \& Schechter]{PreSch74}
        Press, W. H., Schechter, P. 1974, ApJ, 187, 425 [PS]
 
\bibitem[Rees \& Sciama]{ReeSci68}
       Rees, M. J. \& Sciama, D. N. 1968, Nature, 519, 611 

\bibitem[Scoccimarro et al.]{Scoetal00}
        Scoccimarro, R., Sheth, R., Hui, L. \& Jain, B. 2000, ApJ,
546, 20

\bibitem[Seljak]{Sel00}
        Seljak, U. 2000, MNRAS, 318, 203
 
%\bibitem[Seljak \& Zaldarriaga]{SelZal96}
%       Seljak, U., \& Zaldarriaga, M. 1996, ApJ, 469, 437
 
\bibitem[Seljak \& Zaldarriaga]{SelZal00}
        Seljak, U., \& Zaldarriaga, M. 2000, ApJ, 538, 57
 
%\bibitem[Sheth \& Tormen]{SheTor99}
%        Sheth, R. K., \& Tormen, B. 1999, MNRAS, 308, 119 [ST]
 
%\bibitem[Sheth \& Lemson]{SheLem99}
%        Sheth, R. K., \& Lemson, G. 1999, MNRAS, 304, 767
 
\bibitem[Sheth \& Diaferio]{SheDia01}
	Sheth, R. K. \& Diaferio, A. 2001, MNRAS, 322, 901

\bibitem[Springel et al.]{Spretal01}
        Springel, V., White, M., Hernquist, L. 2001, ApJ, 549, 681

\bibitem[Sugerman et al]{Sugetal00}
	Sugerman, B., Summers, F. J. \& Kamionkowski, M. 2000, MNRAS,
311, 762



\bibitem[Sunyaev \& Zel'dovich]{SunZel80}
        Sunyaev, R.A. \& Zel'dovich, Ya. B. 1980, MNRAS, 190, 413
 
%\bibitem[Suto et al.]{Sutetal98}
%        Suto, Y., Sasaki, S., Makino, N. 1998, ApJ, 509, 544
 
 
 
%\bibitem[Tyson \& Angel]{TysAng00}
%                Tyson, A., Angel, R. 2000, {\it The Large-Aperture
%Synoptic Survey Telescope}, in ``New Era of Wide-Field Astronomy'',
%ASP Conference Series.
 
%\bibitem[Van Waerbeke et al]{Vanetal00}
%        Van Waerbeke, L., Mellier, Y., Erben, T. et al. 2000a, A\&A
%submitted, astro-ph/0002500

\bibitem[van den Bosch et al]{vdBetal02}
	van den Bosch F. C., et al, ApJ submitted, astro-ph/0201095.

 
 
\bibitem[Viana \& Liddle]{ViaLid99}
        Viana, P. T. P., Liddle, A. R. 1999, MNRAS, 303, 535

\bibitem[Vishniac]{Vis87}
        Vishniac, E. T., 1987, \ApJ, 322, 597 


\bibitem[Vitvitska et al.]{Vitetal01}
        Vitvitska, M., Klypin, A. A., Kravtsov, A. V. et al. 2001,
preprint, astro-ph/0105349

\bibitem[Weil et al]{Weietal98} 
	Weil, M. L., Eke, V. R., Efstathiou, G. 1998, MNRAS, 300, 773




\bibitem[White]{Whi84}
	White, S. D. M.,  1984, ApJ, 286, 38 

\bibitem[White \& Rees]{WhiRee78}
		White, S. D. M. \& Rees, M. J. 1978, MNRAS, 183, 341


\bibitem[Zurek et al]{Zuretal88}
	Zurek, W. H.,   Quinn, P. J.,  \& Salmon, J. K.  1988, ApJ, 330, 519


\end{thebibliography}
\end{document}